\documentclass[pra, reprint]{revtex4-2}
\usepackage{graphicx} 
\usepackage{dcolumn}  
\usepackage{bm}    
\usepackage{amssymb}  
\usepackage{amsfonts, amsmath, amssymb, mathtools}
\usepackage{amsmath}

\DeclareMathOperator*{\heaviside}{heaviside}
\usepackage{lipsum}
\usepackage[normalem]{ulem}
\usepackage{commath}
\usepackage{xcolor}
\usepackage{comment}
\usepackage[toc,page]{appendix}
\usepackage{booktabs}
\usepackage[normalem]{ulem}

\usepackage[binary-units=true]{siunitx}
\usepackage{hyperref}
\usepackage{cleveref}
\RequirePackage{textcase}

\begin{document}

\raggedbottom

\title{
Enhanced Qubit Readout via Reinforcement Learning}

\author{Aniket Chatterjee}
\email[Corresponding author: ]{aniket.chatterjee@chch.ox.ac.uk}
\affiliation{Department of Physics, University of Oxford}
\affiliation{Centre for Quantum Technologies, National University of Singapore}
\author{Jonathan Schwinger}
\affiliation{Centre for Quantum Technologies, National University of Singapore}
\author{Yvonne Y. Gao}
\email[Corresponding author: ]{yvonne.gao@nus.edu.sg}
\affiliation{Centre for Quantum Technologies,}
\affiliation{Department of Physics, National University of Singapore, Singapore}
\date{\today}

\date{\today}

\begin{abstract}
Measurement is an essential component of robust and practical quantum computation. For superconducting qubits, the measurement process involves the effective manipulation of the joint qubit-resonator dynamics, and it should ideally provide the highest quality for qubit state discrimination with the shortest readout pulse and resonator reset time. Here, we harness model-free reinforcement learning (RL), together with a tailored training environment, to achieve this multifaceted optimization task. Using the IBM quantum device, we demonstrate that the pulse obtained by the RL agent not only successfully achieves state-of-the-art performance, with an assignment error of $(4.6 \pm 0.4)\times10^{-3}$, but also executes the readout and the subsequent resonator reset almost three times faster than the system's default process. Furthermore, the learned waveforms are robust against realistic parameter drifts and follow an intuitive form, making them readily implementable on existing hardware with little computational overhead. Our results provide an effective readout strategy to boost the performance of superconducting quantum processors and demonstrate the value of RL in providing optimal and practical solutions for complex quantum information processing tasks. 
\end{abstract}

\maketitle
\section{Introduction}
Efficient and accurate measurements are critical building blocks for quantum information processing. In circuit quantum electrodynamics (cQED) devices, information from qubits is typically extracted by mapping their states to a resonator, which is strongly coupled to a transmission line. This simple readout mechanism is capable of achieving high-fidelity single-shot state discrimination in typical cQED systems; however, with the ongoing enhancements in both the scale and robustness of current superconducting systems--with gate fidelities now exceeding 99.5\%~\cite{IBMQ} on sizable devices--even small inefficiencies in the readout become a considerable limitation on the overall system performance. As we strive toward practical quantum computation, rapid and effective readout is essential for many critical tasks, such as adaptive quantum error correction protocols~\cite{sivak2023realtime, sudaresan2023_demonstrating} or dynamic quantum circuits that demand many mid-circuit measurements~\cite{corcoles2021_exploiting}.Thus, optimizing the readout process to consistently maximize its efficacy continues to drive active and diverse efforts within the cQED community. 

Remarkable progress has been made regarding hardware improvements, such as enhancing the coherence properties of qubits~\cite{Place2021_new, Wang2022_towards}, creating robust quantum-limited parametric amplifiers~\cite{Jose2020_superconducting, Vijay2011_observation, Eom2012_wideband}, and developing clever integration control capabilities~\cite{Reed2010_fast, Walter2017, Sunada2022_fast, Dassonneville2020, swiadek2024_enhancing}. In parallel, many techniques have also been explored to augment the readout process on existing devices by incorporating optimal integration weights~\cite{Gambetta2007,Ryan2015}, accelerating resonator reset~\cite{Bultink2016, McClure2016_2016}, or applying model-based \textit{in situ} optimization of readout parameters~\cite{Bengtsson2024}.

\begin{figure*}[bht!]
\includegraphics[width=\textwidth]{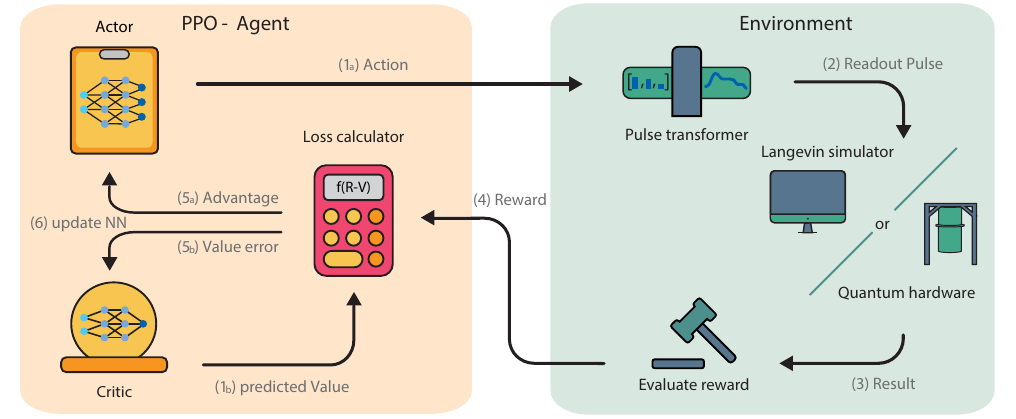}
\caption{
\label{fig:concept}
Depiction of the learning process of a proximal policy optimization (PPO) agent interacting with our environment. Our RL-based readout-optimization framework consists of two main elements: the PPO agent and the environment. In each step of the learning process, the PPO agent samples a batch of actions ($1_\text{a}$) from the policy provided by the actor network. Concurrently, the critic network predicts the value ($1_\text{b}$) of the sampled actions. These actions are processed into the readout pulse (2) through a series of transformations, programmed to reflect the realistic constraints in hardware--i.e., smoothing of the edges to account for limited bandwidth in control electronic--and then tested in a Langevin simulator constructed with real device parameters, emulating the response of a real qubit-resonator system. This simulation's result (3) is then used to calculate the reward (4). The loss calculator then computes the advantage ($5_\text{a}$) and the value error ($5_\text{b}$) by comparing the critic's evaluation against the reward. Finally, the actor and critic neural networks are updated (6) based on the advantage and value error losses, respectively. Our PPO agent iteratively produces increasingly more optimized readout waveforms using this process. }
\end{figure*}

Here, we harness the power of deep reinforcement learning~\cite{Sutton1998, NaturalPolicyGradient, IntroToDeepRL, TRPO} (RL) to optimize the full readout pulse holistically and demonstrate its state-of-the-art performance on a standard superconducting system. More specifically, we construct a physically motivated RL agent capable of rapidly finding waveforms that significantly outperform a default square pulse and offer noticeable improvements against other optimization strategies, such as the cavity-level excitation and reset (CLEAR) pulse~\cite{McClure2016_2016} in fidelity and reset time. We demonstrate the performance of these optimized readout pulses by executing them on IBMQ~\cite{IBMQ} machines. We find that the readout pulses discovered through deep RL achieve overall assignment errors of $(4.6 \pm 0.4) \times 10^{-3}$, on par with the state of the art, while affording a threefold reduction in the time needed to reach maximum fidelity and then reset the resonator back to vacuum compared to the default IBM configurations. Remarkably, in the typical parameter regimes of IBMQ cloud devices, the agent consistently converges on a physically meaningful waveform that can be associated with an analytical form. We dub this the active four-tone readout (A4R), and we show that it can be tuned up through a series of simple calibration protocols that allow direct adaptation to any existing cQED hardware. Finally, we demonstrate that the optimized pulses are robust against system parameter drifts. Our work demonstrates that by leveraging the power of RL, we can effectively enhance the quality and efficiency of transmon qubit readout. Our strategy is, by design, adapted to hardware constraints and easily implementable on generic superconducting devices, making it a powerful tool for quantum information processing with cQED systems. All source code and  parameters are provided on GitHub~\cite{anikenc_2024} for easy reproducibility and adaptation.

The process to read out the state of a superconducting qubit via a resonator can be described by the coherent quantum langevin equation~\cite{cqedReview}, with
\begin{equation}\label{Eq:LangevinEquation}
    \dot{\alpha}_{g/e}(t) = -\left(\frac{\kappa}{2} \mp i\chi\right)\alpha_{g/e}(t) - iA(t),
\end{equation}
where $\alpha_{g/e}$ is the coherent complex-valued resonator state corresponding to the qubit in $|g\rangle$ and $|e\rangle$, respectively, $\dot{\alpha}_{g/e}(t)$ is the time-derivative of the resonator state, $\kappa$ is the energy decay rate, $\chi$ is the qubit-resonator dispersive coupling strength, and $A(t)$ is the coherent input field to the resonator. 

Naively, a longer and more energetic pulse should lead to the largest phase-space separation between the trajectories corresponding to $|g\rangle$ and $|e\rangle$, allowing the most effective state discrimination; however, the dynamics on real hardware are more subtle. Increasing the pulse duration also increases the likelihood of qubit decay due to its finite energy relaxation time. In addition, increases in the pulse amplitude and, correspondingly, the photon population in the resonator, can also worsen qubit coherence due to ionization~\cite{TransmonIonization}. Thus, we must recognize that improving the overall readout performance is a complex, multifaceted optimization process.  In previous studies, various strategies have been developed to improve a specific figure of merit of the readout process in isolation. One of the most notable techniques in this framework is CLEAR~\cite{McClure2016_2016}. This introduces two rectangular segments of fixed duration to the standard square readout pulse to accelerate the resonator ring-up and reset process for devices in the regimes of $\kappa\approx\chi$. In addition, other techniques, such as those investigated in Refs.~\cite{Bultink2016, Hoffer2021_superconducting, Lienhard2021_machine}, have also shown effective improvements via targeted optimization of the ring-up or reset time individually.  

In this study, we use RL to develop a generalized qubit readout process for standard cQED devices that can rapidly achieve the best state discrimination without perturbing the qubit state. This translates into an optimization of the microwave waveform that requires the shortest time and the lowest possible number of photons in the resonator to achieve the highest assignment fidelity, defined as
\begin{equation}
    \mathcal{F} = 1 - \frac{P(g|e) + P(e|g)}{2},
\end{equation}
where $P(g|e)$ $[P(e|g)]$ represents the probability of measuring state $|g\rangle$ ($|e\rangle$) when state $|e\rangle$ ($|g\rangle$) was prepared. Concurrently, we also target an accelerated resonator reset, which enables more rapid repetition of measurements with minimal downtime in between.

\section{Leverage reinforcement learning for readout optimization}

To achieve the target readout performance, we employ a model-free deep RL~\cite{Sutton1998, IntroToDeepRL} framework that seeks out the best waveform that optimizes all relevant figures of merit associated with the readout concurrently and effectively accounts for known system constraints. A conceptual illustration of our implementation is shown in Fig.~\ref{fig:concept}. Compared to other numerical algorithm--such as Nelder-Mead~\cite{NeldMead65} and evolutionary strategies~\cite{PSOPaper}, which have been used in classical-quantum optimization problems~\cite{Evolution_Comparison_Paper}--deep RL uses neural networks to achieve more efficient optimization for complex and multiparameter quantum processes. Using an effective decision-making agent for a reward-oriented problem, RL has been successfully applied to implement real-time quantum error correction~\cite{sivak2023realtime}, for gate calibration~\cite{DRLforGateCal}, and to discover new decoders~\cite{NovelDecoders}. Given the strength of neural networks in learning arbitrary optimal function mappings with little model-based input, a model-free RL optimization framework is particularly well suited to the multipronged readout process. 

\subsection{RL implementation}

We choose the proximal policy optimization (PPO)~\cite{PPOPaper} algorithm due to its strong convergence properties for computationally inexpensive environments. PPO is an actor-critic method~\cite{konda2000ActorCritic}, where a policy neural network (actor) learns optimal actions for a given reward function and a value network (critic) provides a predicted estimate of the reward for the policy. PPO simultaneously ensures good exploration of the action space via the actor and stable learning via the critic. The workflow of the PPO agent is described in Fig.~\ref{fig:concept}. Concretely, for each update of the learning process, the actor outputs the means and standard deviations of a multivariate normal distribution from which an action is sampled. The critic estimates the value of the actions sampled from this action distribution~\cite{Sutton1998} (Fig.~\ref{fig:concept}). During training, the critic network aims to minimize the value error, while the actor aims to produce actions that produce higher rewards than the value estimates of the critic. Training is performed until the critic network accurately estimates the value of the actions and the actor cannot obtain a higher reward than the value estimates, corresponding to finding optimal actions. For the readout problem, this corresponds to estimating the discretized waveform that most effectively maximizes the reward. 

The neural-network training can be implemented using either direct measurement data or realistic simulations of the readout dynamics. Here, due to constraints of cloud access, we opted to perform the training via simulations conducted using the quasiclassical Langevin equations (Eq.~\ref{Eq:LangevinEquation}), which efficiently capture the full dynamics of a resonator coupled to a transmon during a dispersive readout without any costly Hamiltonian simulations. As a result, our method is computationally inexpensive and allows us to obtain all key metrics accurately without simulating high-dimensional Hilbert spaces. With this, we extract the phase-space separation of the resonator state and compute the assignment fidelity, the maximum number of photons, and the residual photon population in the resonator after a reset.

One notable and advantageous feature of PPO is the clipped updates, which corresponds to updating the actor and critic networks with clipped losses, i.e. clipping the loss attributed to the value error and advantage to be between $-\epsilon$ and $+\epsilon$, where $\epsilon$ is a small hyperparameter. This prevents overshooting, whereby the networks might fall into some local minima of reward with suboptimal actions, and it also allows for better convergence to reduce the computational overhead incurred by the learning process. Further details about the implementation of the algorithm and the hyperparameters used are presented in Appendix~A.

\begin{figure}[t!]
\includegraphics[width=\columnwidth]{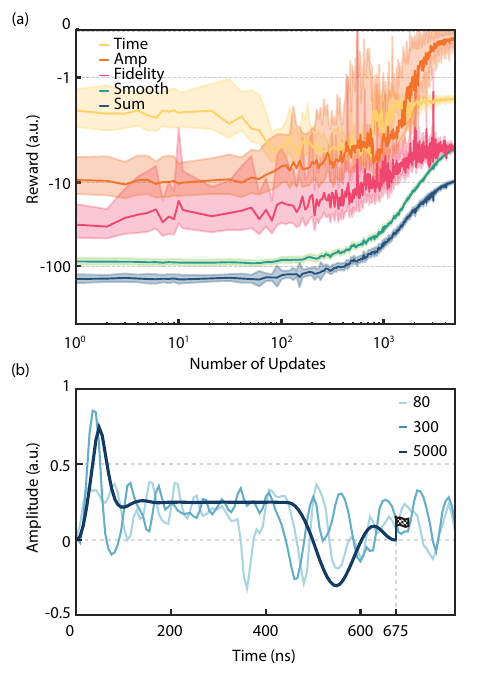}
\caption{
\label{fig:learning}
RL learning curves and resulting waveforms for Brisbane. (a) Learning curves showing the evolution of the total reward, and individual reward contributions over a training run of 5000 updates. The shaded regions show the standard deviations within the batch during training. Initially, fidelity optimization is prioritized while the time reward decreases slightly. Between 1000 and 5000 updates, the fidelity converges to the maximum value allowed by the physical system, while the reset time and auxiliary rewards, such as smoothness and reset amplitudes, are further optimized until convergence. (b) Intermediate waveforms after the action of the pulse transformer at 80, 300, and 5000 updates during training. The initial waveforms are all seeded from randomly generated pulses. Within 300 updates, a distinct high-amplitude ring-up within the limit of the maximum photon population emerges. The agent makes significant improvements between 300 and 5000 updates to smooth the waveform and achieve a fast reset, within 675\,ns.}
\end{figure}

\begin{figure*}[htbp]
\includegraphics[]{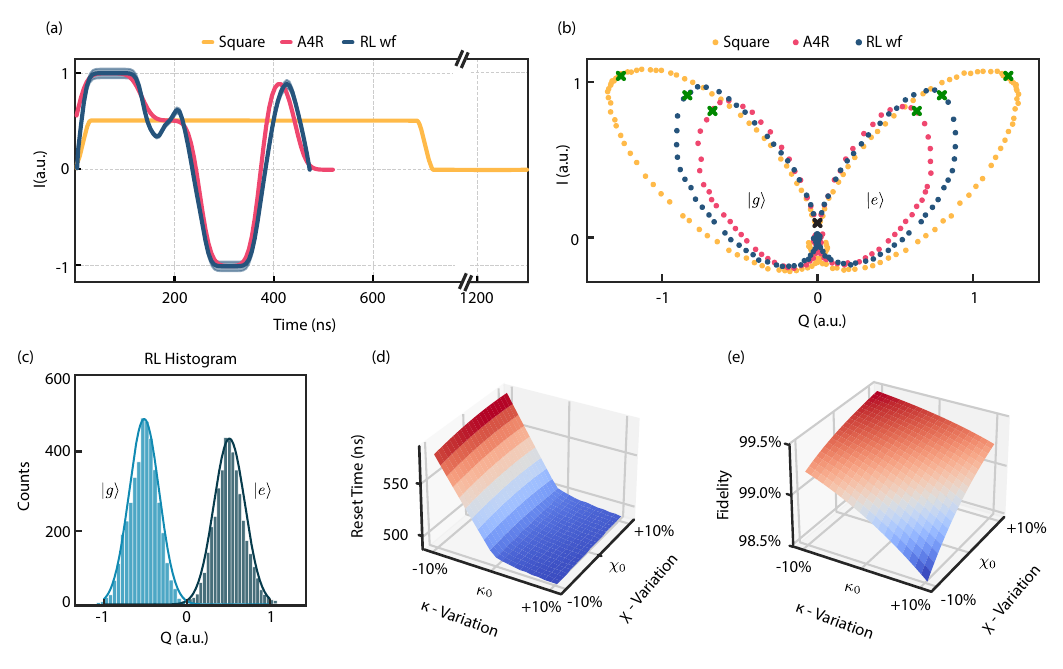}
\caption{
\label{fig:waveforms}
Performance and stability of optimized waveforms discovered via RL. (a) An exemplary waveform (blue) for IBM Kyoto discovered by our physically informed RL agent. Its smooth amplitude modulation and low bandwidth allow direct implementation on hardware. Both the learned waveform and its generalized analytical counterpart, A4R (red), lead to low assignment errors of $(4.6 \pm 0.4)\times 10^{-3}$ and $(5.7 \pm 0.8)\times 10^{-3}$ and reset the resonator within 470\,ns, significantly faster than the 1400\,ns needed using the default readout configuration in Kyoto. We use calibrated integration weights around the time of maximum assignment fidelity at 264\,ns for A4R and the RL waveform. More details are included in Appendix~C. (b) Measured phase-space trajectories of the coherent states in the readout resonator corresponding to the qubit in states $|g\rangle$ and $|e\rangle$, achieved with the default readout pulse (yellow), RL optimized pulse (blue), and A4R (red), respectively. Notably, the RL and A4R trajectories achieve maximum fidelities at much smaller separations than the square pulse, demonstrating that shorter readouts, which do not necessarily reach steady state, can be more optimal. The black cross marks the beginning of the readout and the green crosses indicate the end of the drive to populate the readout resonator and the start of the reset to vacuum. (c) An exemplary readout histogram from 4096 single-shot measurements on IBM Kyoto using the waveform discovered by our RL agent, showing effective state discrimination with $>5\sigma$ separation between the $|g\rangle$ and $|e\rangle$ distributions. Simulated stability landscapes for the (d) reset time and (e) assignment fidelity under $\kappa$ and $\chi$ variations of up to $10\%$. Both show that these key figures of merit do not fluctuate significantly under realistic parameter drifts.
}
\end{figure*}

\subsection{Reward function}
The crux of our optimization process involves implementing the appropriate reward function, which is crucial for improving learning and optimization efficiency~\cite{RL_Rewards}, as well as physical viability. All of the relevant optimization objectives and constraints described above are integrated into our reward function:
\begin{align}\label{Eq:Reward Function}
    R = &-k_1 \log_{10}(1 - \max(\mathcal{F}(t))) - k_2 \kappa \tau_R \nonumber \\ 
       &- k_3 \int (A''(t))^2 dt - k_4 (A(0) + A(t_{min,N})) \\ 
       &- k_5 \mathrm{ReLU}(\max(N(t)) - N_0) \nonumber \\ 
       &- k_6 \heaviside(t_{max,\mathcal{F}} - t_{min,N}),\nonumber
\end{align}
where $k_i$ are the individual coefficients associated with each reward term; their specific values are detailed in Appendix~B. 

First, we maximize $\mathcal{F}$, which is computed from the phase-space trajectories~\cite{cqedReview}. We use a term proportional to the logarithm of the assignment error to steadily increase the reward over a wide range of assignment errors. This motivates the agent to focus on obtaining a maximum in $\mathcal{F}$ before optimizing other auxiliary objectives. Next, we minimize the quantity $\kappa \tau_R$, where $\tau_R$ is the total reset time. To ensure practical feasibility on real devices, we account for a conservative bandwidth of 50\,MHz on the control system's digital-to-analog converter output and encourage smooth pulses by minimizing $\int (A''(t))^2 dt$ $[$where $A''(t)$ is the second derivative of the waveform$]$ and penalizing the terminal amplitudes $A(0)$ and $A(t_{min, N})$ at the start and reset point of the readout. Finally, the photon population $N(t)$ is limited to be below that of the default square readout ($N_0$) on the test device. A final order penalty term ensures that the RL agent always achieves high fidelity and resets the resonator in the correct order.

\section{Implementation on IBMQ Devices}

We test the efficacy of our RL-based optimization strategy on IBMQ systems via cloud access. We chose two devices, Kyoto and Brisbane, which cover the range of typical system parameters used in superconducting qubit processors. 

We obtain relevant system parameters to construct an appropriate learning environment through a series of standard calibration measurements. Specifically, we characterize the dispersive coupling strength $\chi$ between the qubit and the readout resonator as well as the decay rate of the resonator $\kappa$. For the results reported in this work, we choose two devices: ibm\_brisbane Qubit 2 (Brisbane) with $\kappa = 21.4$\,MHz and ${2\chi}/{2\pi} = 0.31$\,MHz, operating in the regime $\kappa/\chi \approx 22 $, and ibm\_kyoto Qubit 2 (Kyoto) with $\kappa = 10.4$\,MHz, ${2\chi}/{2\pi} = 0.85$\,MHz, and $\kappa/\chi\approx 4$. Both devices have a default measurement time of 1400\,ns, consisting of a 720-ns square readout pulse and a 680-ns delay for passive resonator reset. The resulting assignment errors using these default square pulses are $(5.8 \pm 0.9)\times10^{-3}$ and $(7.0 \pm 1.0)\times10^{-3}$, respectively.

Testing our RL-enabled readout strategy on these two devices covers the range of typical system parameters used in superconducting qubit readout, with Brisbane representing the fast-readout configuration $\kappa\gg\chi$ made possible through the inclusion of Purcell filters~\cite{PurcellFilter}, while Kyoto operates at the more familiar $\kappa\sim\chi$ point.

\section{Results and discussions}
\subsection{Learning outcomes}
Using the parameters obtained from our calibration measurements on these two devices, we train the PPO agent with batch sizes of 128 steps for 5000 updates. This procedure is implemented on an Nvidia Tesla P100 graphics processing unit with a wall-clock training time of approximately 3 min, which indicates a very efficient learning process and makes it a practical yet powerful strategy for readout optimization. The learning curve for the Brisbane environment is shown in Fig.~\ref{fig:learning}(a), where the normalized sum reward and its dominant contributions due to assignment fidelity, reset amplitude, pulse smoothness, and total reset time are shown as functions of the learning updates. We visualize the intermediate learned policies by plotting the mean action across the batch at the 80th, 300th, and 5000th updates in Fig.~\ref{fig:learning}(b) to gain insights into the agent's progression during this process. 

Initially, the agent is completely stochastic and produces random waveforms; however, within the first 80 updates, the agent develops features corresponding to the ring-up, readout, and reset processes. Initially, the agent prioritizes maximizing fidelity, leading to a slightly higher amplitude ring-up in the 80th update waveform and a flat readout segment to reach the appropriate fidelity. From the 80th to 300th update, the agent iteratively improves the reset time, as shown by the higher-amplitude ring-up tone, along with the notable formation of two reset tones for a high-quality reset. From the 300th update onward until the 5000th update, the agent improves the reset amplitude deviation and smooths the waveform till we obtain a final converged waveform at the 5000th update. This is a clear demonstration that the agent approaches the task holistically, balancing the different figures of merit as well as practical constraints of an effective readout throughout the learning process to achieve the optimal solution. 

Remarkably, the sequences learned by our RL agent for both Brisbane and Kyoto consistently follow a well-defined structure consisting of four distinct segments. We dub this the active four-tone readout (A4R). A4R follows a simple analytical form, with a single high-amplitude ring-up tone, a constant calibrated segment, and a two-tone reset. A detailed description of the parameters governing A4R is presented in Appendix~D. As an illustration, we show the default square pulse, the learned pulse obtained by the RL agent, and the corresponding A4R waveform in Fig.~\ref{fig:waveforms}(a). In addition, we show that the action of the optimized pulses (RL and A4R) follows the intuitive trajectories in phase space (Fig.~\ref{fig:waveforms}(b)), starting with a sharp increase in the I component, followed by a constant-amplitude drive which leads to maximum separation in the Q component, without necessarily driving the resonator to its steady state. Finally, the active reset is performed to effectively reduce the final photon population to below $0.05$, similar to the residual population associated with the default readout on the test device. Although the first reset tone can clear the resonator state from $>50$ to $<1$ photon, a short second kickback tone is required to ensure that the small fraction of remaining excitations is rapidly evacuated. This two-tone reset, with segments of variable durations and amplitudes, ensures a faster, more flexible, and more complete reset of the resonator compared to the CLEAR sequence, which requires the two reset tones to be of equal duration~\cite{McClure2016_2016}. A more detailed comparison with CLEAR is presented in Appendix~E. The key practical advantage of the A4R protocol is the generality of its dynamics. By modeling it as a photon-transfer process, we can construct a set of simple calibration measurements, as detailed in Appendix~D, to obtain the optimal A4R parameters without further numerical optimization.

\subsection{Performance of optimized waveforms}
We execute the resulting optimized waveforms on IBM Kyoto (Q2) and Brisbane (Q2) and verify their performance on IBMQ devices via cloud access. The default readout configurations for these qubits consist of a square pulse and a passive set, with total length 1400\,ns and maximum numbers of photons of 26 and 59, respectively. In comparison, the RL agent achieves equal or slightly superior fidelities ($\approx 99.5\%$) using the same number of photons but offers significantly shorter measurement and reset times, with total durations of 675 and 470\,ns, respectively. An exemplary phase-space trajectory of the resonator state is shown in Fig.~\ref{fig:waveforms}(b). This shows that both the RL pulse and A4R follow intuitive dynamics to achieve effective state discrimination (Fig.~\ref{fig:waveforms}(c)) and realize effective state reset at the end of the process. A summary of all relevant figures of merit for the readout performance is provided in Appendix~C. 

Moreover, the resulting assignment fidelities also exhibit strong stability against realistic system-parameter drifts~\cite{DriftPaper}. We test this by simulating the readout properties over a range of $\kappa$ and $\chi$ that mimics a $\pm 10\%$ fluctuation on the real hardware. The landscapes of the resulting performance metrics are shown in Fig.~\ref{fig:waveforms}(d) and (e). The assignment fidelity varies by $\leq1\%$ for $\kappa/\chi$ drifts within $\pm10\,\%$, while the reset time varies by $<80$\,ns and is only dependent on changes in $\kappa$. 

\subsection{Connection to other optimization techniques}
These demonstrations show that we have constructed a physically informed RL agent to effectively optimize the readout process of a superconducting qubit with minimal computational overhead. We implemented the learned protocol obtained by the agent on IBMQ devices and showed that they successfully achieve high-fidelity readout with accelerated resonator reset, offering a significant reduction in the total measurement duration. 

Importantly, the optimized sequences exhibit well-defined intuitive structures and are robust against hardware parameter drifts, highlighting their practical applicability to real quantum processors. The convergence to the generalized A4R sequence closely resembles the analytical pulses such as CLEAR~\cite{McClure2016_2016} and that in Ref.~\cite{Hazra2025}. A more quantitative comparison with CLEAR is provided in Appendix~E. This feature indicates that our framework is capable of discovering physically meaningful sequences without any predefined models, making it a versatile tool that can be used to first seek out optimal waveforms when the physical model of the system is not well understood. Once a stable solution is found, as in the case of A4R here, it can be generalized into a few-parameter numerical optimization for easy implementation on real hardware.

Furthermore, the stable structure of the resulting waveform across different system parameters also makes it a versatile tool to augment other hardware-based readout-optimization approaches. For instance, in Ref.~\cite{Spring2024}, the speedup is attained by tailoring the readout and filter mode on the device such that an extremely fast readout (i.e. a vastly enhanced $\kappa$) can be performed without degrading the qubit coherence. In another recent work~\cite{swiadek2024_enhancing}, a speedup of the readout process was achieved through physically manipulating the spectral separations of the qubit and readout resonator such that the effective $\chi$ was enhanced during the readout. With the ability to achieve a stable and robust readout performance across different $\kappa/\chi$ regimes, our technique can be readily adapted to work in conjunction with these hardware upgrades to further enhance the overall readout performance.

Overall, our work harnesses the intersectional knowledge between model-free reinforcement learning and quantum information science to attain tangible performance enhancement in superconducting readout. Our work also serves to demonstrate the general efficacy and versatility of RL in complex optimization problems for practical quantum computing applications.

\section{Acknowledgement} 
We thank Mr. Arthur Strauss, Dr. Lirande Pira, Mr. Beng Yee Gan and Dr. Patrick Rebentrost for the insightful discussions and feedback on this work. We acknowledge support from Horizon Quantum Computing in providing access to the IBM cloud devices, and we thank Mr. Kyle Timonthy Chu for his assistance in executing test demonstrations. We acknowledge funding from the National Research Foundation (NRFF12-2020-0063) and the Ministry of Education (MOE-21-0054-P0001), Singapore. 

\section*{Appendix A: Reinforcement Learning Implementation}\label{appendix:rl}
For the readout-optimization task, we use policy-based reinforcement learning. Broadly speaking, the algorithm tries to find the optimal policy function $\pi_{\theta}(s|a)$ that chooses the best action $a$ to take at a given state $s$, parametrized by a neural network with weights $\theta$. In this context, $s$ does not refer to the quantum state, but instead represents all current information describing the environment, e.g., the system parameters and final qubit state. Thus, $s$ is simply a constant vector, as the system does not change during the training process and is reset after each step of the optimization.

More concretely, we use the proximal policy optimization (PPO), which is a widely used policy-based algorithm known for its robust learning and convergence capabilities without expansive memory requirements. PPO is an actor-critic method, in which the actor network (policy) outputs the actions to take at a given state, and the critic network tries to predict the value of the actions taken. The actor tries to maximize its advantage $A = R - V$, where $R$ is the reward for a given action and $V$ is the value predicted by the critic. The critic focuses on improving its target estimates $(V - R)^2$. As the actor, through the learning process, increases its likelihood of producing high-advantage actions, the critic improves its estimates of the actor's reward.

Here, we implement the PPO algorithm in \textsc{JAX}~\cite{jax2018github}, based on \textsc{PUREJAXRL}~\cite{lu2022discovered}. We opt for shared hidden layers~\cite{shengyi2022the37implementation} for the actor and critic to reduce the training complexity and for faster learning. In addition, we perform batch rollouts~\cite{Batched_Paper} of simulations to speed up training. The hyperparameters were optimized using a grid search, and the results of this search are listed in Table~\ref{table:hyper}.

\begin{table}[htbp]
\centering
\begin{tabular}{ p{4.5cm} p{2.5cm} }
    \toprule\toprule
    \textbf{Hyperparameter}           & \textbf{Value} \\
    \midrule
    Learning rate & $0.0003$ \\
    Hidden layer size     & 128\\
    Hidden layers     & 2\\
    No. vectorized environments                  & 128\\
    Activation function       & ReLU6 \\
    Optimizer                 & Adam  \\ 
    Update epochs               & 4  \\ 
    No. minibatches                 & 4  \\ 
    Clip epsilon                 & 0.2  \\ 
    Value clip epsilon                 & 0.2  \\ 
    Entropy coefficient                 & 0  \\  
    Maximum normed gradient                 & 0.5  \\
    \bottomrule
\end{tabular} 
\caption{Optimal hyperparameters used for the PPO algorithm. The same set of parameters is used to obtain the waveforms for both Kyoto and Brisbane. }
\label{table:hyper} 
\end{table}

\section*{Appendix B: Readout Environment}\label{appendix:env}
The readout environment takes a 1D real vector of pulse amplitudes as input, which corresponds to the readout waveform sent to the resonator. As the default readout pulse is approximately 720ns long on IBM Quantum devices, we discretize the RL readout pulse to take 121 real amplitudes, corresponding to a sampling time of 6\,ns. Real-valued pulses are sufficient, as we focus on systems with low self-Kerr and operate within $N << N_c$ where $N_c$ is the critical photon population~\cite{CriticalPhotonPaper}. We clip the pulse amplitudes to $±\mu$, where $\mu$ is the maximum amplitude available on the IBMQ system relative to the default amplitude used for measurement. We also smooth the waveform using a Gaussian convolution to mimic the low-pass filter response of a typical arbitrary waveform generator. Finally, we scale up the waveform by the steady-state amplitude $A_0=0.5\sqrt{N_0(\kappa^2 + 4\chi^2)}$, where $N_0$ is the default measurement photon population. We obtain the values of $N_0$ and the resonator energy decay rate $\kappa$ through ac Stark shift measurements~\cite{schuster2005_ac} on each of the test devices. 

\begin{table*}[hbt]\label{table:ro_performance}
\begin{ruledtabular}
\begin{tabular}{c|ccccc}
\textbf{Readout configuration} &  Duration (ns) & 1 - $\mathcal{F}_{max}$ & $t_{\mathcal{F}_{max}}$ (ns) & $N_{max}$ & $N_{min}$  \\ \hline
\textbf{Kyoto default + delay} &  1400 & $(5.8 \pm 0.9)\times10^{-3}$  & 560 & 26.8 & $<0.05$ \\
\hline
\textbf{RL} & \textbf{470}  & $(\textbf{4.6} \pm \textbf{0.4})\times10^{-3}$  & \textbf{264} & \textbf{26.2} & $(0.04 \pm 0.03)$ \\
\hline
\textbf{A4R} & 490 & $(5.7 \pm 0.8)\times10^{-3}$  & \textbf{264} & 26.5 & $(0.02 \pm 0.02$ \\
\hline
\hline
\textbf{Brisbane default + delay}\,\, &  1400 & $(\textbf{7.0} \pm \textbf{1.0})\times10^{-3}$  & 733 & \textbf{58.6} & $<0.05$ \\
\hline
\textbf{RL} & \textbf{675}  & $(\textbf{7.0} \pm \textbf{1.0})\times10^{-3}$  & \textbf{542} & 59.1 & $(0.05 \pm 0.03)$ \\
\hline
\textbf{A4R} & 685 & $(7.2 \pm 1.0)\times10^{-3}$  & \textbf{542} & 59.2 & $(0.03 \pm 0.03)$ \\
\end{tabular}
\end{ruledtabular}
\caption{Comparison of readout performances between the default measurement protocol, the RL-optimized waveform, and a calibrated A4R waveform on Kyoto and Brisbane. Bold numbers correspond to the best achieved value in the category. On both systems, the RL and A4R waveforms show similar improvements of a two- to threefold speedup over the default pulse with similar or even lower assignment errors. This is attributed to the high amplitude ring-up causing the resonator to achieve minimum error more rapidly. Furthermore, the reset for both devices reduces the resonator photon population to $<0.05$, comparable with the default configurations used on these test devices. 
\label{tab:readout performance}}
\end{table*}

The processed waveform is passed to the Langevin simulation, described by Eq.~\ref{Eq:LangevinEquation}. From the simulated coherent state amplitude, we obtain the photon population $N(t)=|\alpha(t)|^2$ and the state separation $S(t)=|\alpha_+(t) - \alpha_-(t)|$. To calculate the resultant time-dependent fidelity, we model the resonator as a Gaussian state and assume we integrate for a short boxcar window that can acquire the instantaneous signal. Hence, the extracted fidelity from the state separation in phase space is given by
\begin{equation}\label{Eq:SNR}
    F_{\mathrm{SNR}}(t) = 0.5 \left(1 + \text{erf}\left(\lambda S(t)\right)\right),
\end{equation}
where $\lambda$ is a scaling constant that captures amplification from a quantum-limited amplifier and assumes constant noise for the distribution. The state-separation fidelity $F_{\mathrm{SNR}}$ represents the fidelity of the measurement if we assume perfect state preparation and no qubit evolution during the readout. We account for the infidelity associated with qubit decay and measurement-induced state transitions by adding a qubit decay rate that increases linearly with the resonator photon population~\cite{MeasDecay_Paper}. This qubit fidelity term, $F_q$, is given by
\begin{equation}
    F_q(t) = \exp\left(-\gamma_0t - \gamma_P \int_0^t N(\tau)d\tau\right).
\end{equation}
Finally, the overall fidelity is given by
\begin{equation}\label{eq:fidelity_fit}
    \mathcal{F}(t)=0.5\left(1 + \text{erf}\left(F_0 \times \lambda S(t) \times F_q(t)\right)\right),
\end{equation}
where $F_0$ is the qubit initialization fidelity. Notably, $F_q$ decays with time, while the state separation $S(t)$ increases with time. Thus, it is clear that there is an optimal measurement duration, and this is not necessarily the steady-state duration due to realistic qubit decay. Finally, we evaluate $\max{\mathcal{F}(t)}$, which is then used in the reward function. To capture the realistic behavior of the hardware, we extract these parameters from the chosen devices by fitting Eq.~\ref{eq:fidelity_fit} to the measured assignment infidelities obtained with various different acquisition times. 

To calculate the readout duration, we assume that the measurement ends at the first photon minimum after maximum fidelity is achieved, as enforced by the order penalty in the reward function. The reset time of the readout corresponds to $t_{min, N}$ along with the additional time required to reach the target photon population $N_I$. We calculate this by assuming standard exponential decay of the photon population. Thus, the total reset time $\tau_R$ is
\begin{equation}
    \tau_R = t_{min, N} + \frac{m}{\kappa} \times \log\left(\frac{N(t_{min, N})}{N_I}\right),
\end{equation}
where $m$ is a $T_1$ scaling factor added to discourage the agent from conducting a partial reset and then waiting. We find $m = 8$ is sufficient to force the agent to actively reset all systems regardless of $\kappa$.

The reward function for the environment is given by Eq.~\ref{Eq:Reward Function}, as detailed in the main text. We choose reward coefficients $k_i$ to prioritize fidelity, after which the auxiliary objectives of faster readout and smoothing are included. Hence, we set the fidelity coefficient $k_1 = 10$, the time coefficient $k_2 = 2$, and the smoothness coefficient $k_3 = 1$. The time and smoothness coefficients are determined through grid sweeps of training runs, and provided they are significantly smaller than the fidelity coefficient, we recover the waveforms presented in the main text. We set penalty-term coefficients, such as the amplitude coefficient $k_4$ and the photon coefficient $k_5$, to be large values of 100, such that they are enforced throughout the training run.

We use \textsc{DIFFRAX}~\cite{DiffraxPaper} and \textsc{JAX} for the ordinary differential equation simulation to implement the readout environment. These tools allow us to leverage just-in-time compilation and automatic vectorization to execute fast parallel training runs and hyperparameter-optimization sweeps.

\section*{Appendix C: Summary of system parameters and readout performance}\label{appendix:ro_summary}

A comparison of the readout performance on both IBM Kyoto and Brisbane is presented in Table~\ref{tab:readout performance}, and a summary of the readout pulse parameters is provided in Table~\ref{tab:params}. We identify the time of maximum fidelity using repeated fidelity measurements at various acquisition times for each readout. For the infidelities of the RL waveforms and A4R, we integrate the signal for a window centered around the time of maximum fidelity with calibrated integration weights provided by the IBM devices. For the default readout, we report the infidelity as integrated with the custom weights provided by IBM, which achieves the same fidelity within shot noise as integrating for a window around the time of maximum fidelity. We are not able to further optimize the specific form of these integration weights due to restrictions on IBM devices.

\begin{table}[htbp]
\centering
\begin{tabular}{ p{3cm} p{2.5cm} p{2.5cm}}
    \toprule\toprule
    \textbf{Parameters}  & \textbf{Brisbane Q2} & \textbf{Kyoto Q2}\\
    \midrule
    qubit frequency            & 4.610\,GHz      & 4.733\,GHz\\ 
    resonator frequency            & 7.229\,GHz      & 7.225\,GHz\\ 
    $\chi$ to readout    & 0.15\,MHz      & 0.42\,MHz\\
    readout $\kappa$    & 21.4\,MHz        & 10.4\,MHz\\
    qubit $T_1$           & 	291\,$\mu$s     &  344\,$\mu$s\\
    qubit $T_2$                  &  222\,$\mu$s   &  281\,$\mu$s\\
    \bottomrule
\end{tabular} 
\caption{Parameters for the two qubits used in our implementation of the optimized readout pulses. }
\label{tab:params} 
\end{table}

\section*{Appendix D: Generalized Active 4-Tone Readout (A4R)}\label{appendix:a4r}
\begin{figure}[h!]
    \centering
    \includegraphics{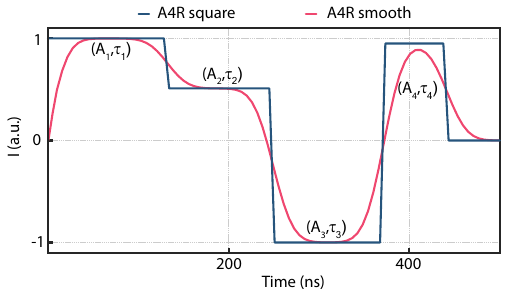}
    \caption{A4R waveform, which consists of four segments: ring up, steady-state readout, reset, and kickback. Each segment is characterized by two variables: duration $\tau_i$ and amplitude $A_i$. The blue line depicts the square pulse form of A4R, while the red line shows the smoothed form after bandwidth constraints are applied. }
    \label{fig:A4R_waveform}
\end{figure}

The A4R pulse is designed to achieve high fidelity and rapid resonator reset in the regime of $\kappa/\chi \gg 1$.  It consists of four sequential segments: ring-up, steady-state readout, photon depletion, and kickback (see Fig.~\ref{fig:A4R_waveform}). Each segment $i$ is characterized by an amplitude $A_i$ and a duration $\tau_i$, resulting in a total of eight parameters that can be extracted efficiently from measurement outcomes or a sample-efficient optimizer such as Nelder-Mead~\cite{NeldMead65}.

In this regime, a single ring-up segment is sufficient, as the phase induced by the dispersive interaction is small. Thus, one calibrated ring-up tone effectively drives the resonator to reach near steady state in phase space, after which the second tone ensures a high-fidelity state discrimination. The amplitude during the steady-state readout segment, $A_2$, is fixed at a target steady-state amplitude, while $A_0$ and the remaining seven parameters can be calibrated using a simple three-step measurement routine.

We set the ring-up amplitude $A_1 = \mu A_0$ and solve the coherent Langevin equation for a constant-amplitude tone to estimate the ring-up duration $\tau_1$ as 
\begin{equation}\label{eq:tau_1}
    \tau_1 = \frac{2}{\kappa}\ln\left(\frac{\mu}{\mu - 1}\right),
\end{equation}
where first-order approximations in $\chi t$ are used to eliminate oscillating terms. As the ring-up should be as quick as possible, we use this estimate of $\tau_1$ to find the largest $A_1$ within the bandwidth constraint. We sweep the estimated $\tau_1$ value and perform photon-number measurements to get the final value of $\tau_1$. Then, $\tau_2$ is calibrated to maximize fidelity by measuring the instantaneous fidelity while varying this duration. 

For resonator reset, two tones are required to account for the impact of $\kappa$ and $\chi$ on the trajectory. To achieve fast depletion of photons in the resonator, we select $A_3=-\mu A_0$. The duration, $\tau_3$, can then be calculated as
\begin{equation}\label{eq:tau_2}
    \tau_3 = \frac{2}{\kappa}\ln\left(\frac{\mu + 1}{\mu}\right),
\end{equation}
which is independent of $\chi$. Any inaccuracies in the reset are corrected by the kickback tone with, the parameters $A_4$ and $\tau_4$ determined directly via the residual photon population measurements. For the Brisbane setup, the first reset tone removes the bulk of the energy, bringing its photon number from 58.8 to 0.6. The final kickback further reduces the population to $<0.05$ photons, comparable to that achieved in the default readout via passive decay of the resonator.

\section*{Appendix E: Comparison with CLEAR}\label{appendix:clear}

In this study, we have demonstrated that A4R provides a significant improvement in the readout efficiency for the test devices in the $\kappa\gg\chi$ regime. Here, we provide an analysis of its performance against CLEAR (Ref.~\cite{McClure2016_2016}), which enacts an active ring-up and reset alongside a standard steady-state square pulse. Focusing on the $\kappa \approx \chi$ regime, CLEAR applies the ansatz of playing two segments of equal duration with variable amplitudes before and after a square pulse. The four optimal amplitudes associated with the ring-up and reset are analytically determined in the linear regime and, more generally, can be obtained numerically with the appropriate target photon populations in mind. For instance, in Ref.~\cite{McClure2016_2016}, the authors used a target of $N_0$ after ring-up and $\approx 0.1$ photons after the reset. Overall, they found a speedup of 50\% in resonator reset compared to using a passive process following a square readout on the specific test device. 

To compare the A4R readout with CLEAR, we reproduce the optimization protocol outlined in Ref.~\cite{McClure2016_2016} to determine CLEAR pulse parameters for the Kyoto device using a Nelder-Mead~\cite{NeldMead65} optimizer implemented from \textsc{SCIPY}~\cite{SciPy}. The resulting optimized parameters for CLEAR show a noticeably faster reset ($\approx 460\,$ns) compared to the default ($\approx 680\,$ns); however, no discernible speedup is found for the ring-up stage in this $\kappa \gg \chi$ regime when a constraint is imposed on the maximum number of photons in the optimization. Overall, for Kyoto, CLEAR requires $\approx 1180$\,ns for a full readout cycle. While this offers an improvement over the default (1400\,ns), it significantly underperforms compared to A4R, which completes the full readout process in 470\,ns under the same optimization constraints. 

Overall, alongside a threefold speedup over the simple square readout with passive decay, A4R demonstrates a 2.5-fold speedup over the CLEAR waveform. While A4R and CLEAR are quite reminiscent of each other structurally, several improvements in A4R--such as a single segment ring-up, optimized readout durations, and flexible reset segment durations for the $\kappa/\chi\gg 1$ regime--allow for significantly improved readout speeds while preserving a robust assignment fidelity.

\begin{table}[htbp]
\centering
\label{table:rl_clear} 
\begin{tabular}{ p{2cm} p{3cm} p{3cm}}
    \toprule\toprule
    \textbf{Device parameter}  & \textbf{RL/A4R} & \textbf{CLEAR}\\
    \midrule
    $\kappa/\chi$  = 0.5     & 99.4\% ; 967\,ns  & 99.4\% ; 1100\,ns \\
    $\kappa/\chi$  = 2     & 99.5\% ; 502\,ns  & 99.5\% ; 757\,ns \\
    $\kappa/\chi$  = 5     & 99.4\% ; 265\,ns  & 99.5\% ; 457\,ns \\
    $\kappa/\chi$  = 10     & 99.6\% ; 164\,ns  & 99.7\% ; 276\,ns \\
    \bottomrule
\end{tabular} 
\caption{Simulated readout fidelity and total readout duration (measurement + reset) for different system parameters using CLEAR and the waveforms discovered by RL. } \label{table: compare_clear}
\end{table}

To assess the performance comparison more extensively, we also simulated the action of CLEAR and A4R over a wide range of $\kappa/\chi$ values. In Table~\ref{table: compare_clear}, we compare both the resulting readout fidelities and the durations of the pulses discovered by the RL agent and the CLEAR pulses derived analytically for the same system parameters. In these simulations, we fix the dispersive coupling strength $\chi$ to match that of Kyoto Qubit 2, and we vary the $\kappa/\chi$ ratio between 0.5, 2, 5, and 10. We set the other relevant system parameters, such as the photon population of the readout $N_0$ and the decay rates $\gamma_0$ and $\gamma_P$, such that a square readout pulse achieves a fidelity of 0.995. In this setting, as there is no reference square readout, we reduce the readout segment of the waveform until the time of maximum fidelity, similar to that in A4R. We note that this in itself represents a significant speedup compared to the CLEAR readout, which previously required a longer duration near steady state. The code for reproducing these results is provided on GitHub~\cite{anikenc_2024}.

This comparison shows that even in regimes where $\kappa$ is large, as pursued in some hardware optimization strategies, the optimized sequence produced by RL still offers a clear acceleration in the total readout and reset process compared to the CLEAR pulse. As such, our framework serves as a valuable and versatile tool that can be applied to augment the performance of superconducting qubit readout across different device parameters or hardware configurations.

\bibliography{references}    
\end{document}